\documentclass[10pt]{article}
\usepackage{graphicx}
\usepackage{amsmath}
\usepackage{amssymb}
\usepackage{caption2}
\begin{document}
\bibliographystyle{prsty}
\begin{center}
{\large {\bf \sc{  Analysis of the $\Sigma_Q$  baryons in the nuclear matter  with   the   QCD sum rules  }}} \\[2mm]
Zhi-Gang Wang \footnote{E-mail,wangzgyiti@yahoo.com.cn.  }   \\
 Department of Physics, North China Electric Power University, Baoding 071003, P. R. China
\end{center}

\begin{abstract}
In this article, we extend our previous work to   study the  $\Sigma$-type
heavy baryons $\Sigma_c$ and $\Sigma_b$ in the nuclear matter using the QCD sum rules, and obtain three coupled QCD sum rules for
 the  masses $M_{\Sigma_Q}^*$, vector self-energies $\Sigma_v$ and pole residues $\lambda^*_{\Sigma_Q}$ in the nuclear matter. Then
 we take into account the effects of the
  unequal pole residues  from different spinor structures, and   normalize the masses
 from the  QCD sum rules in the vacuum to the experimental
  data, and obtain the mass-shifts $\delta M_{\Sigma_c}=-123\,\rm{MeV}$ and $\delta M_{\Sigma_b}=-375\,\rm{MeV}$ in the nuclear matter.
\end{abstract}



\section{Introduction}

The  QCD sum rules is a powerful theoretical tool in studying the
ground state hadrons both in the vacuum and in the nuclear matter, and has given many
successful descriptions of the hadron  properties, such as the masses, decay constants, form-factors,
hadron coupling constants, etc \cite{SVZ79}.
The  properties of the light-flavor mesons and baryons  in the nuclear matter have been studied extensively with the
QCD sum rules \cite{Drukarev1991,C-parameter,C-parameter-2}, while the existing works on  the
 heavy quarkonia  and heavy baryons in the nuclear matter  focus on the $J/\psi$, $\eta_c$, $D$,  $B$, $D_0$, $B_0$, $D^*$,
 $D_1$, $\Lambda_c$ and $\Lambda_b$ \cite{Jpsi-etac,Hayashigaki,WangLambda}.
 The in-medium mass modifications of the  $Q\bar{q}$ and $Qqq$ hadrons  differ greatly
 from the corresponding ones of the $q\bar{q}$ and $qqq$ hadrons due to the appearance of the heavy quark,
 the full propagators of the heavy  quarks in the nuclear matter undergo much slight modifications compared with that of the light quarks.
 The upcoming FAIR (facility for
antiproton and ion research) project at GSI (heavy ion research lab)  provides the opportunity to extend the experimental
studies of the in-medium properties of the mesons and baryons into the charm sector \cite{CBM,PANDA}, we can study the heavy hadrons in nuclear
matter with the QCD sum rules and make predictions to be confronted with the experimental data of the CBM (compressed baryonic matter)
 and $\rm{\bar{P}ANDA}$ collaborations in the future.

 The scattering amplitude of one-gluon exchange can be rephrased into an antisymmetric  antitriplet  $\overline{\bf{3}}_{ c}$
and an symmetric sextet $\bf{6}_{ c}$ in the color-space. The attractive interaction
in the  antisymmetric  antitriplet favors the formation of the diquark states in the color
 antitriplet  $\overline{\bf{3}}_{ c}$,   the most stable diquark states, the $\Lambda$-type diquark states, maybe exist
in  the color antitriplet $\overline{\bf{3}}_{ c}$, flavor
antitriplet $\overline{\bf{3}}_{ f}$ and spin singlet ${\bf{1}}_s$
channels due to Fermi-Dirac statistics \cite{Color-Spin}, and the next stable diquark states, the $\Sigma$-type diquark states, maybe exist
in  the color antitriplet $\overline{\bf{3}}_{ c}$, flavor
sextet ${\bf{6}}_{ f}$ and spin triplet ${\bf{3}}_s$
channels \cite{WangHdiquark}. In the heavy quark limit, the heavy baryons can be classified  as the $\Lambda$-type or  $\Sigma$-type baryons according to the
spin structures of the two light quarks \cite{ReviewH}.  In Ref.\cite{WangLambda}, we study the $\Lambda$-type heavy baryons $\Lambda_c$ and $\Lambda_b$
 in the nuclear matter using the QCD sum rules, and obtain the in-medium positive mass-shifts
  $\delta M_{\Lambda_c}=51\,\rm{MeV}$ and $\delta M_{\Lambda_b}=60\,\rm{MeV}$, respectively.
  In this article, we extend our previous work  to study the  properties of the $\Sigma$-type heavy baryons $\Sigma_c$ and $\Sigma_b$ in the nuclear
matter using the QCD sum rules.

The article is arranged as follows:  we study the
   heavy baryons $\Sigma_c$ and $\Sigma_b$ in the nuclear matter
 with  the  QCD sum rules in Sec.2; in Sec.3, we present the
numerical results and discussions; and Sec.4 is reserved for our
conclusions.

\section{The in-medium $\Sigma_Q$ baryons  with  QCD sum rules}
We study the  $\Sigma_c$ and $\Sigma_b$ baryons  in the nuclear matter
with the following two-point correlation functions $\Pi(q)$,
\begin{eqnarray}
\Pi(q) &=& i\int d^4 x\, e^{iq\cdot x}\langle \Psi_0| T\left\{J(x)\bar{J}(0)\right\}| \Psi_0\rangle \, , \nonumber \\
J(x)&=&\epsilon^{ijk} u^t_i(x)C\gamma_\mu d_j(x)\gamma^\mu \gamma_5 Q_k(x)\, ,
\end{eqnarray}
where the $i$, $j$, $k$ are color indexes, $Q=c,b$, the $C$ is the charge conjunction matrix,
and  the $ |\Psi_0\rangle$ is the nuclear matter ground state.
The correlation functions  $\Pi(q)$ can  be decomposed as
\begin{eqnarray}
\Pi(q)&=& \Pi_s(q^2,q \cdot u)+ \Pi_q(q^2,q \cdot u) \!\not\!{q} +\Pi_u(q^2,q \cdot u)  \!\not\!{u}\,,
\end{eqnarray}
according to Lorentz covariance,  parity and time reversal invariance \cite{Drukarev1991,C-parameter}.
In  the limit $u_\mu=(1,0)$,  the component $\Pi_i(q^2,q \cdot u)$ reduces  to $\Pi_i(q_0,\vec{q})$,  where $i=s,q,u$.

We  insert a complete set  of intermediate heavy baryon states with the
same quantum numbers as the current operators $J(x)$ into the correlation functions $\Pi(p)$ to obtain the hadronic representation
\cite{SVZ79}, then isolate the ground state $\Sigma_Q$ baryon contributions, and resort to  the dispersion relation to rephrase  the
three components of the correlation functions  $\Pi_i(q_0,\vec{q})$ in the following form:
\begin{eqnarray}
\Pi_i(q_0,\vec{q})&=&{1\over2\pi i}\int _{-\infty}^\infty~d\omega{\Delta\Pi_i(
\omega,\vec{q})\over\omega-q_0} \, ,
\end{eqnarray}
where
\begin{eqnarray}
\Delta\Pi_s(\omega,\vec{q})&=&-2\pi i \frac{\lambda_{\Sigma_Q}^{*2}M_{\Sigma_Q}^*}{2E^*_q}\left[
\delta(\omega-E_q)-\delta(\omega-\bar{E}_q)\right]\,,\nonumber\\
\Delta\Pi_q(\omega,\vec{q})&=&-2\pi i\frac{\lambda_{\Sigma_Q}^{*2} }{2E^*_q}\left[\delta(\omega-E_q)
-\delta(\omega-\bar{E}_q)\right]\,,
\nonumber\\
\Delta\Pi_u(\omega,\vec{q})&=&+2\pi i\frac{\lambda_{\Sigma_Q}^{*2}\Sigma_v }{2E^*_q}\left[\delta
(\omega-E_q)
-\delta(\omega-\bar{E}_q)\right]\,,
 \end{eqnarray}
 $E_q^*= \sqrt{ M_{\Sigma_Q}^{*2}+\vec{q}^2}$,
$E_q=\Sigma_v+E_q^*$, $\bar{E}_q=\Sigma_v-E_q^*$,   the $M_{\Sigma_Q}^{*}$, $\Sigma_v$ and $\lambda^*_{\Sigma_Q}$ are
 the  masses, vector self-energies and pole residues of the $\Sigma_Q$ baryons  respectively in the nuclear matter.

We carry out the operator product expansion in the nuclear matter  at the large space-like region $q^2\ll 0$, and obtain the
  spectral densities at the level of quark-gluon degrees of freedom, then
take the limit $u_\mu=(1,0)$, and write the three components $\Pi_i(q_0,\vec{q})$  in the following form \cite{Drukarev1991,C-parameter}:
\begin{eqnarray}
\Pi_i(q_0,\vec{q})&=&\sum_n C_n^i(q_0,\vec{q})\langle{O}_n\rangle_{\rho_N}\, ,
\end{eqnarray}
 where the $C_n^i(q_0,\vec{q})$ are the Wilson coefficients,  the $\langle{O}_n\rangle_{\rho_N}$, which are defined as $\langle \Psi_0|{O}_n|\Psi_0\rangle$,
  are the condensates in the nuclear matter and
   can be decomposed as $\langle{\cal{O}}\rangle+\rho_N\langle
{\cal{O}}\rangle_N$  at the low nuclear density, the   $\langle{\cal{O}}\rangle$ and $\langle
{\cal{O}}\rangle_N$ denote the vacuum condensates and the nuclear matter induced condensates,  respectively. The imaginary parts of the QCD spectral densities
can be obtained  through the formula $\Delta\Pi_i(\omega,\vec{q})={\rm{limit}}_{\epsilon\to 0} \left[\Pi_i(\omega+i\epsilon,\vec{q})-\Pi_i(\omega-i\epsilon,\vec{q}) \right]$.

Finally, we  match  the hadronic spectral densities with the QCD  spectral densities,
and multiply both sides with  the weight function $(\omega-\bar{E}_q)e^{-\frac{\omega^2}{M^2}}$,    perform
the integral  $\int_{-\omega_0}^{\omega_0}d\omega$,
\begin{eqnarray}
\int_{-\omega_0}^{\omega_0}d\omega\Delta\Pi_i(\omega,\vec{q})(\omega-\bar{E}_q)e^{-\frac{\omega^2}{M^2}}\,,
\end{eqnarray}
to exclude  the  negative-energy pole contributions,
  and obtain the following three QCD sum rules:
\begin{eqnarray}
\lambda_{\Sigma_Q}^{*2}e^{-\frac{E_q^2}{M^2}} &=&\int_{m_Q^2}^{s_0^*}ds\int_{x_i}^1 dx \left\{ \frac{x(1-x)^3(s-\widehat{E}_Q^2)(5s-3\widehat{E}_Q^2)}{32\pi^4}-\frac{2x(1-x)\langle q^{\dagger}iD_0q\rangle_{\rho_N}}{3\pi^2}\right. \nonumber\\
&&\left[2+(\widetilde{m}_Q^2-2s)\delta(s-\widehat{E}_Q^2)\right]+\frac{1}{96\pi^2}\langle\frac{\alpha_sGG}{\pi}\rangle_{\rho_N}\left[ (4-5x) +(1-x)\widetilde{m}_Q^2\delta(s-\widehat{E}_Q^2)\right] \nonumber\\
&&-\frac{m_Q^2}{144\pi^2}\langle\frac{\alpha_sGG}{\pi}\rangle_{\rho_N}\frac{(1-x)^3}{x^2}\left( 1+\frac{\widetilde{m}_Q^2}{2M^2}\right)\delta(s-\widehat{E}_Q^2)+\bar{E}_q\left[  \frac{x(1-x)\langle q^\dagger q\rangle_{\rho_N}}{2\pi^2} \right.\nonumber\\
&& -\frac{x\langle q^{\dagger} g_s \sigma G q\rangle_{\rho_N}}{6\pi^2}\delta(s-\widehat{E}_Q^2) +\frac{x(1-x)\langle q^{\dagger} g_s \sigma G q\rangle_{\rho_N}}{12\pi^2}\left(1+\frac{2s}{M^2}\right)\delta(s-\widehat{E}_Q^2)  \nonumber\\
&&-\frac{x\langle q^\dagger iD_0iD_0 q\rangle_{\rho_N}}{3\pi^2}\delta(s-\widehat{E}_Q^2)
-\frac{x(1-x)\langle q^\dagger iD_0iD_0 q\rangle_{\rho_N}}{\pi^2}\left(3+\frac{4\widetilde{m}_Q^2-6s}{3M^2}\right)\delta(s-\widehat{E}_Q^2)\nonumber\\
&&\left.\left.-\frac{x\langle q^{\dagger} g_s \sigma G q\rangle_{\rho_N}}{24\pi^2}\delta(s-\widehat{E}_Q^2)
+\frac{(1-x)\langle q^{\dagger} g_s\sigma G q\rangle_{\rho_N}}{8\pi^2} \delta(s-\widehat{E}_Q^2)
\right]\right\}e^{-\frac{s}{M^2}}+\frac{\langle\bar{q}q\rangle^2_{\rho_N}}{3}e^{-\frac{E_Q^2}{M^2}}\, , \nonumber \\
\end{eqnarray}

\begin{eqnarray}
\frac{\lambda_{\Sigma_Q}^{*2}M_{\Sigma_Q}^*}{m_Q}e^{-\frac{E_q^2}{M^2}} &=&\int_{m_Q^2}^{s_0^*}ds\int_{x_i}^1 dx \left\{ \frac{3(1-x)^2(s-\widehat{E}_Q^2)^2}{64\pi^4} -\frac{\langle q^{\dagger}iD_0q\rangle_{\rho_N}}{3\pi^2}+\frac{2(1-x)\langle q^{\dagger}iD_0q\rangle_{\rho_N}}{3\pi^2}\right. \nonumber\\
&&\left[1+2s\delta(s-\widehat{E}_Q^2) \right] -\frac{(1-x)^2m_Q^2}{192\pi^2x^3} \langle\frac{\alpha_sGG}{\pi}\rangle_{\rho_N}\delta(s-\widehat{E}_Q^2)+\frac{1-2x^2}{64\pi^2x^2}\langle\frac{\alpha_sGG}{\pi}\rangle_{\rho_N} \nonumber\\
&&+\bar{E}_q\left[  \frac{(1-x)\langle q^\dagger q\rangle_{\rho_N}}{2\pi^2} -\frac{\langle q^{\dagger}g_s \sigma G q\rangle_{\rho_N}}{6\pi^2}\delta(s-\widehat{E}_Q^2)+\frac{(1-x)\langle q^{\dagger}g_s \sigma G q\rangle_{\rho_N}}{4\pi^2}\delta(s-\widehat{E}_Q^2)
 \right. \nonumber\\
&&-\frac{(1-x)\langle q^{\dagger}g_s \sigma G q\rangle_{\rho_N}}{6\pi^2}\left(1-\frac{s}{M^2}\right)\delta(s-\widehat{E}_Q^2)-\frac{\langle q^\dagger iD_0iD_0 q\rangle_{\rho_N}}{\pi^2}\delta(s-\widehat{E}_Q^2)
\nonumber\\
&&+\frac{3(1-x)\langle q^\dagger iD_0iD_0 q\rangle_{\rho_N}}{\pi^2}\delta(s-\widehat{E}_Q^2)-\frac{2(1-x)\langle q^\dagger iD_0iD_0 q\rangle_{\rho_N}}{\pi^2}\left(
1-\frac{s}{M^2}\right)\delta(s-\widehat{E}_Q^2)\nonumber\\
&&\left.\left.-\frac{(2x-1)\langle q^{\dagger}g_s \sigma G q\rangle_{\rho_N}}{8\pi^2x}\delta(s-\widehat{E}_Q^2)
\right]\right\}e^{-\frac{s}{M^2}} +\frac{2\langle\bar{q}q\rangle^2_{\rho_N}+\langle q^\dagger q\rangle^2_{\rho_N}}{3}e^{-\frac{E_Q^2}{M^2}}\, ,
\end{eqnarray}

\begin{eqnarray}
\lambda_{\Sigma_Q}^{*2}\Sigma_v e^{-\frac{E_q^2}{M^2}} &=&\int_{m_Q^2}^{s_0^*}ds\int_{x_i}^1 dx \left\{ \frac{x(1-x)(7s-5\widehat{E}_Q^2)\langle q^{\dagger} q\rangle_{\rho_N}}{4\pi^2}  -\frac{x\langle q^{\dagger}g_s\sigma G q\rangle}{12\pi^2} \left[5+2\widetilde{m}_Q^2\delta(s-\widehat{E}_Q^2) \right]
\right. \nonumber\\
&&-\frac{x(1-x)\langle q^{\dagger}g_s \sigma G q\rangle_{\rho_N}}{2\pi^2}\left[-\frac{3}{4}+\left(\frac{\widetilde{m}_Q^2}{6}-\frac{3s}{2}-\frac{s \widetilde{m}_Q^2}{3M^2}\right)\delta(s-\widehat{E}_Q^2) \right] \nonumber\\
&&-\frac{x\langle q^{\dagger}iD_0iD_0q\rangle_{\rho_N}}{6\pi^2}\left[5+2\widetilde{m}_Q^2\delta(s-\widehat{E}_Q^2) \right]  +\frac{x(1-x)\langle q^{\dagger}iD_0iD_0q\rangle_{\rho_N}}{\pi^2}\nonumber\\
&& \left[\frac{5}{2}+\left( -\frac{5\widetilde{m}_Q^2}{3}+9s+\frac{2s\widetilde{m}_Q^2}{M^2}\right)\delta(s-\widehat{E}_Q^2)\right]+\frac{x\langle q^{\dagger}g_s\sigma G q\rangle}{8\pi^2} \left[\frac{1}{2}+\frac{\widetilde{m}_Q^2}{3}\delta(s-\widehat{E}_Q^2) \right] \nonumber\\
&&\left.+\frac{(1-x)\langle q^{\dagger}g_s\sigma G q\rangle}{16\pi^2} +\bar{E}_q  \frac{4x(1-x)\langle q^{\dagger}iD_0 q\rangle_{\rho_N}}{3\pi^2}
\left[4+\widetilde{m}_Q^2\delta(s-\widehat{E}_Q^2) \right]\right\}e^{-\frac{s}{M^2}}\nonumber\\
&&+\bar{E}_q \frac{2\langle q^\dagger q\rangle^2_{\rho_N}}{3}e^{-\frac{E_Q^2}{M^2}}\, ,
\end{eqnarray}
where $\widetilde{E}_Q^2=\frac{m_Q^2}{x}+\vec{q}^2$, $E_Q^2=m_Q^2+\vec{q}^2$, $x_i=\frac{m_Q^2}{s}$, $s^*_0=\omega_0^2=s_0-\vec{q}^2$,
 the $\omega_0$ is the threshold parameter, and
  $x_i\rightarrow 0$  in the spectral densities where the function $\delta(s-\widehat{E}_Q^2)$ appears.
We can obtain the  masses $M^*_{\Sigma_Q}$, vector
self-energies $\Sigma_v$ and pole residues $\lambda^*_{\Sigma_Q}$ in the nuclear matter by solving the three
equations with simultaneous  iterations.

\section{Numerical results and discussions}
 The input parameters of the QCD sum rules in the nuclear matter are taken as
 $\langle q^\dagger q\rangle_{\rho_N}={3\over2}\rho_N$,
 $\langle \bar{q} q\rangle_{\rho_N}=\langle \bar{q} q\rangle+{\sigma_N\over
m_u+m_d}\rho_N $, $\langle\bar{q}q\rangle=(-0.23\,\rm{GeV})^3$,  $m_u+m_d=12\,\rm{MeV}$,
$\sigma_N=45\,\rm{MeV}$,
$\langle\frac{\alpha_sGG}{\pi}\rangle_{\rho_N}=\langle\frac{\alpha_sGG}{\pi}\rangle-0.65\,{\rm GeV}\rho_N$,
$\langle\frac{\alpha_sGG}{\pi}\rangle=(0.33\,\rm{GeV})^4$,
$\langle q^{\dagger} iD_0iD_0 q\rangle_{\rho_N}+{1 \over 12}\langle q^{\dagger}g_s\sigma G q\rangle_{\rho_N}=0.031\,{\rm{GeV}}^2\rho_N$,
$\langle \bar{q} iD_0iD_0 q\rangle_{\rho_N}+{1\over 8}\langle\bar{q}g_s\sigma G q\rangle_{\rho_N}=0.3\,{\rm{GeV}}^2\rho_N$,
$\langle\bar{q}g_s\sigma G q\rangle_{\rho_N}=\langle\bar{q}g_s\sigma G q\rangle+3.0\,{\rm GeV}^2\rho_N$,
$\langle q^{\dagger}g_s\sigma G q\rangle_{\rho_N}=-0.33\,{\rm GeV}^2\rho_N$, $\langle q^\dagger iD_0 q\rangle_{\rho_N}=0.18\,{\rm GeV}\rho_N$,
$\langle\bar{q}g_s\sigma Gq\rangle=m_0^2\langle\bar{q}q\rangle$,  $m_0^2=0.8\,\rm{GeV}^2$,
$\langle \bar{q}q\rangle^2_{\rho_N}=f\langle \bar{q}q\rangle_{\rho_N}\times\langle \bar{q}q\rangle_{\rho_N}+(1-f)\langle \bar{q}q\rangle \times \langle \bar{q}q\rangle$, $f=0.5\pm0.5$, $\vec{q}^2=(0.27\,\rm{GeV})^2$, and
$\rho_N=(0.11\,\rm{GeV})^3$ \cite{C-parameter,C-parameter-2}.

We  recover the QCD sum rules in the vacuum by taking the limit  $\rho_N=0$,  then differentiate  the Eqs.(7-8) with respect to  $\frac{1}{M^2}$ respectively,
and eliminate the pole residues $\lambda_{\Sigma_Q}$ (here we smear the star $*$ to denote the pole residues in the vacuum),
then obtain  two QCD sum rules for  the masses  $M_{\Sigma_Q}$, one can consult Ref.\cite{WangzgEPJA} for the technical details.

 In the conventional QCD sum rules \cite{SVZ79}, there are two
criteria (pole dominance and convergence of the operator product
expansion) for choosing  the Borel parameter $M^2$ and threshold
parameter $s_0$.  We impose the two criteria on the heavy baryon
states  $\Sigma_Q$,  and choose the threshold parameters $s_0=(10.0\pm0.5)\,\rm{GeV}$ and $(43.5\pm0.5)\,\rm{GeV}$,   the Borel parameters
 $M^2=(1.9-2.7)\,\rm{GeV}$  and $(4.8-5.6)\,\rm{GeV}$ for the heavy baryons $\Sigma_c$  and $\Sigma_b$, respectively \cite{WangzgEPJA}.
 Finally we  obtain the values
of the masses and pole resides  $M_{\Sigma_c}=(2.54\pm0.15)\,\rm{GeV}$ and $(2.42\pm0.20)\,\rm{GeV}$,  $M_{\Sigma_b}=(5.96\pm0.10)\,\rm{GeV}$ and $(5.73\pm0.16)\,\rm{GeV}$, $\lambda_{\Sigma_c}=(5.4\pm 1.4)\times 10^{-2}\,\rm{GeV}^3$ and $(3.6\pm 1.0)\times 10^{-2}\,\rm{GeV}^3$,   $\lambda_{\Sigma_b}=(8.0\pm 1.7)\times 10^{-2}\,\rm{GeV}^3$ and $(5.0\pm 1.2)\times 10^{-2}\,\rm{GeV}^3$ from the QCD sum rules with respect to the spinor structures  $\!\not\!{q}$ and $1$, respectively \cite{WangzgEPJA}.

The experimental values of the masses are $M_{\Sigma_c^{++}}=(2454.03\pm0.18)\,\rm{MeV}$, $M_{\Sigma_c^+}=(2452.9\pm0.4)\,\rm{MeV}$, $M_{\Sigma_c^0}=(2453.76\pm0.18)\,\rm{MeV}$,  $M_{\Sigma_b^+}=(5807.8\pm2.7)\,\rm{MeV}$ and  $M_{\Sigma_b^-}=(5815.2\pm2.0)\,\rm{MeV}$  from the Particle Data Group \cite{PDG}, the average values are
$M_{\Sigma_c}=2.454\,\rm{GeV}$ and $M_{\Sigma_b}=5.812\,\rm{GeV}$, respectively, which are consistent  with the new experimental data from the CDF  collaboration \cite{CDF}. The predicted masses from both spinor
structures $\!\not\!{q}$ and $1$ can reproduce the experimental data approximatively, however, the pole residues from the spinor structures  $\!\not\!{q}$ and $1$
differ from each other greatly, i.e. the values have the hierarchy $\lambda^{\gamma \cdot q}_{\Sigma_Q} \gg \lambda_{\Sigma_Q}^{1}$, where the upper indexes denote the spinor structures. We can draw the conclusion that the two QCD sum rules  in Eqs.(7-8) in the limit $\rho_N=0$ can be satisfied by the approximate equal masses but unequal pole residues.
If we obtain the masses $M_{\Sigma_Q}$  by dividing Eq.(8) with Eq.(7) in the limit $\rho_N=0$,
the predictions $M_{\Sigma_c}=1.40^{+0.08}_{-0.05}\,\rm{GeV}$ and  $M_{\Sigma_b}=3.56^{+0.14}_{-0.10}\,\rm{GeV}$ are much smaller than the experimental  data due to the unequal pole residues.  We can multiply the smaller masses by some coefficients to offset the effects of the unequal pole residues and reproduce the experimental data.

Take  the same Borel parameters and threshold parameters as the QCD sum rules in the vacuum \cite{WangzgEPJA},  we can obtain the hadronic parameters $M^*_{\Sigma_c}=1.33^{+0.06}_{-0.03}\,\rm{GeV}$, $M^*_{\Sigma_b}=3.33^{+0.09}_{-0.07}\,\rm{GeV}$, $\lambda_{\Sigma_c}^*=2.46^{+0.22}_{-0.16}\times 10^{-2}\,\rm{GeV}^3$, $\lambda_{\Sigma_b}^*=1.25^{+0.08}_{-0.04}\times 10^{-2}\,\rm{GeV}^3$, $\Sigma_{v}^{\Sigma_c}=0.446^{+0.035}_{-0.027}\,\rm{GeV}$,
  $\Sigma_{v}^{\Sigma_b}=0.776^{+0.042}_{-0.035}\,\rm{GeV}$ from the three coupled QCD sum rules in Eqs.(7-9). In the limit $\rho_N= 0$, we can obtain the values $M_{\Sigma_c}=1.40^{+0.08}_{-0.05}\,\rm{GeV}$,   $M_{\Sigma_b}=3.56^{+0.14}_{-0.10}\,\rm{GeV}$, $\lambda_{\Sigma_c}=1.99^{+0.29}_{-0.26}\times 10^{-2}\,\rm{GeV}^3$, $\lambda_{\Sigma_b}=8.73^{+0.90}_{-0.65}\times 10^{-3}\,\rm{GeV}^3$. In calculations, we have taken the assumption that the pole residues $\lambda^*_{\Sigma_Q}$ in the QCD sum rules (see Eqs.(7-9)) have the same values. In fact,  those QCD sum rules can be satisfied with approximately the same in-medium masses $M^*_{\Sigma_Q}$ and vector self-energies  $\Sigma^{\Sigma_Q}_v$,  but different pole residues $\lambda^*_{\Sigma_Q}$, which can be denoted as $\lambda^*_{\Sigma_Q}(1)$, $\lambda^*_{\Sigma_Q}(2)$ and $\lambda^*_{\Sigma_Q}(3)$ from the QCD sum rules in Eqs.(7-9) respectively. We draw this conclusion tentatively based  on the QCD sum rules for the heavy baryon states in the vacuum \cite{WangzgEPJA},
  and normalize the masses from the QCD sum rules in the vacuum to the experimental data to  study the mass modifications in the nuclear matter,
   \begin{eqnarray}
   \delta M_{\Sigma_Q}&=&\frac{M^*_{\Sigma_Q}-M_{\Sigma_Q}}{M_{\Sigma_Q}}\times M_{\Sigma_Q}^{\rm{exp}}\, ,
   \end{eqnarray}
and obtain the central values $\delta M_{\Sigma_c}=-123\,\rm{MeV}$ and $\delta M_{\Sigma_b}=-375\,\rm{MeV}$, the $\delta M_{\Sigma_Q}$ are the scalar self-energies  $\Sigma_s^{\Sigma_Q}$.
The $\Sigma_Q$ baryons have a heavy quark besides two light quarks,
  the heavy quark  interacts  with the nuclear matter through  the exchange of the intermediate gluons, the contributions from the gluon condensates are
  of minor importance and the modifications of the gluon condensates in the nuclear matter are mild,   we  expect
 the ratios of the mass-shifts $\delta M_{\Sigma_Q}/M_{\Sigma_Q}$  are smaller than that of the nucleons.
From Fig.1, we can obtain the mass differences $\Delta M_{\Sigma_Q}=M^*_{\Sigma_Q}-M_{\Sigma_Q}$, $\Delta M_{\Sigma_c}=-0.07\pm 0.02\,\rm{GeV}$, $\Delta M_{\Sigma_b}=-0.24\pm 0.04\,\rm{GeV}$, the uncertainties of the mass differences originate from the Borel parameters  are about $29\%$ and $17\%$, respectively. The ratios are $\frac{\Delta M_{\Sigma_c}}{M_{\Sigma_c}}=-5.0\%$  and $\frac{\Delta M_{\Sigma_b}}{M_{\Sigma_b}}=-6.7\%$,  the mass modifications are rather  small.

  If we take into account the uncertainties of the heavy quark masses  and threshold parameters,  $\delta m_c=\pm0.1\,\rm{GeV}$, $\delta m_b=\pm0.1\,\rm{GeV}$, $\delta s^0_{\Sigma_c}=\pm0.5\,\rm{GeV}^2$,  $\delta s^0_{\Sigma_b}=\pm0.5\,\rm{GeV}^2$ \cite{WangzgEPJA}, the values of the mass differences  $\Delta M_{\Sigma_Q}$ survive approximately, no additional uncertainties are introduced.
 At the interval $f=0\sim1$, the masses $M^*_{\Sigma_Q}$ decrease monotonously
 with the increase of the parameter $f$, the uncertainty $\delta f=\pm 0.5$   leads  to the uncertainties
   $\pm0.06\,\rm{GeV}$ and  $\pm 0.21\,\rm{GeV}$ for the  masses $M^*_{\Sigma_c}$ and $M^*_{\Sigma_b}$ respectively in the nuclear matter,  and the corresponding  uncertainties  for the mass-shifts $\delta M_{\Sigma_c}$  and $\delta M_{\Sigma_b}$ are $\pm 105\,\rm{MeV}$ and $\pm 343\,\rm{MeV}$,  respectively.

If we take the Ioffe current to interpolate the proton,  the QCD sum rules indicate that there exists  a positive
vector self-energy  $\Sigma^N_v=(0.23-0.35)\,\rm{GeV}$ with the typical values of the relevant condensates
and other input parameters,  which is consistent with
the values of the vector self-energies in the relativistic nuclear physics phenomenology, on the other hand, although
 the scalar self-energy depends strongly on the in-medium  four-quark condensate and the nucleon $\sigma$ term,
a reasonable negative scalar self-energy can be obtained with the suitable parameters \cite{C-parameter}.
There exists significant cancelation between the scalar and vector self-energies,
which leads to a quasinucleon energy close to the free-space nucleon mass and satisfies
 the empirical expectation  that the quasinucleon energy is shifted
only slightly in nuclear matter relative to the free-space mass.
The in-medium self-energies $\Sigma^N_s$ and $\Sigma^N_v$ can be written as
 \begin{eqnarray}
 \Sigma^N_s&=&-\frac{8\pi^2\sigma_N\rho_N}{ M^2(m_u+m_d)} \, ,\nonumber\\
\Sigma^N_v&=&\frac{32\pi^2\rho_N}{M^2} \, ,
 \end{eqnarray}
in the leading  order approximation,  ${\Sigma^N_s}/{\Sigma^N_v}\approx -1$, which  indicates  a
substantial cancelation between  self-energies $\Sigma^N_s$ and $\Sigma^N_v$ in the nuclear matter.
The mean-field models  predicate  that  the typical self-energies of the nucleons  in nuclear matter saturation density are  $\Sigma^N_s\approx -350 \,\rm{ MeV}$ and $\Sigma^N_v\approx +300 \,\rm{MeV}$ respectively, which  correspond to the real energy-independent optical potentials $S$ and $V$, and  significant cancelation between the potentials occurs,   the effective non-relativistic central
potential is about tens of $\rm{MeV}$. If the same mechanism works for the $\Sigma_c$ baryon, the vector self-energy should be $\Sigma^{\Sigma_c}_v\approx +123\,\rm{MeV}$ rather than $+446\,\rm{MeV}$ according to  the unequal pole residues  $\lambda^*_{\Sigma_Q}(1)\neq \lambda^*_{\Sigma_Q}(2)\neq\lambda^*_{\Sigma_Q}(3)$, the total self-energy  $\Sigma^{\Sigma_c}_s+\Sigma^{\Sigma_c}_v \approx 0$  under the condition  $\lambda^*_{\Sigma_Q}(3)/\lambda^*_{\Sigma_Q}(2)\approx 1.9$, then the quasi-$\Sigma_c$  energy in the nuclear matter close to the free-space $\Sigma_c$ mass.
  The present prediction of the mass-shift $\delta M_{\Sigma_c}=-123\,\rm{MeV}$ can be confronted with the experimental data  from the
   CBM and $\rm{\bar{P}ANDA}$ collaborations in the future \cite{CBM,PANDA}, where
 the  properties of the charmed baryons in the nuclear matter will be studied.

In  the non-relativistic
harmonic-oscillator potential model,  the spectrum of the bound
states (the energies $E_n$ and the wave-functions $\psi_n(x)$) and the
exact correlation functions are known precisely. In Ref.\cite{ChSh3}, Lucha,  Melikhov and Simula try to fit the effective threshold parameter
 so as to reproduce both the ground energy $E_0$
and the pole residue $R_0$ ($\psi^*_0(0)\psi_0(0)$),   or reproduce the ground energy $E_0$ only
and take the pole residue $R_0$ as a calculated parameter. They observe that
the pole residue $R_0$  is determined to a great extent by the continuum contributions, and draw the conclusion
that the ground-state parameters extracted from the QCD sum rules have uncontrolled  systematic errors if the continuum contributions  are not known exactly
 and  modeled by an effective continuum.
In the real QCD world, the hadronic spectral densities are not known
exactly, the pole residues or decay constants  in some (or most) cases  are not experimentally
measurable quantities, and should be calculated by some theoretical
approaches, the true values are difficult to obtain. On the other hand, the hadronic  spectrum densities
  in the non-relativistic
harmonic-oscillator potential model are of the Dirac $\delta$
function type even in the limit $n\to \infty$,  while in the case of the
QCD,  the widths of the higher radial excited states become broader
 gradually before submerging into the asymptotic quarks and gluons.
For example,   the widths of the  $\pi$, $\pi(1300)$, $\pi(1800)$,
$\cdots$ are $\sim 0\,\rm{GeV}$, $(0.2-0.6)\,\rm{GeV}$,
$0.208\pm0.012\,\rm{GeV}$, $\cdots$ respectively, and  the widths
of the $K$, $K(1460)$, $K(1830)$, $\cdots$ are $\sim 0 \,\rm{GeV}$,
$\sim (0.25-0.26)\,\rm{GeV}$, $\sim 0.25\,\rm{GeV}$, $\cdots$
respectively \cite{PDG}. We  cannot
estimate the unknown systematic uncertainties of the QCD sum rules
before the spectral densities in both the QCD and phenomenological
sides are  known with great accuracy. In the present case, the situation is even worse, the theoretical
 works on the $\Sigma_Q$ baryons in the nuclear matter
are rare, and no experimental data  exist.

\begin{figure}
 \centering
 \includegraphics[totalheight=5cm,width=7cm]{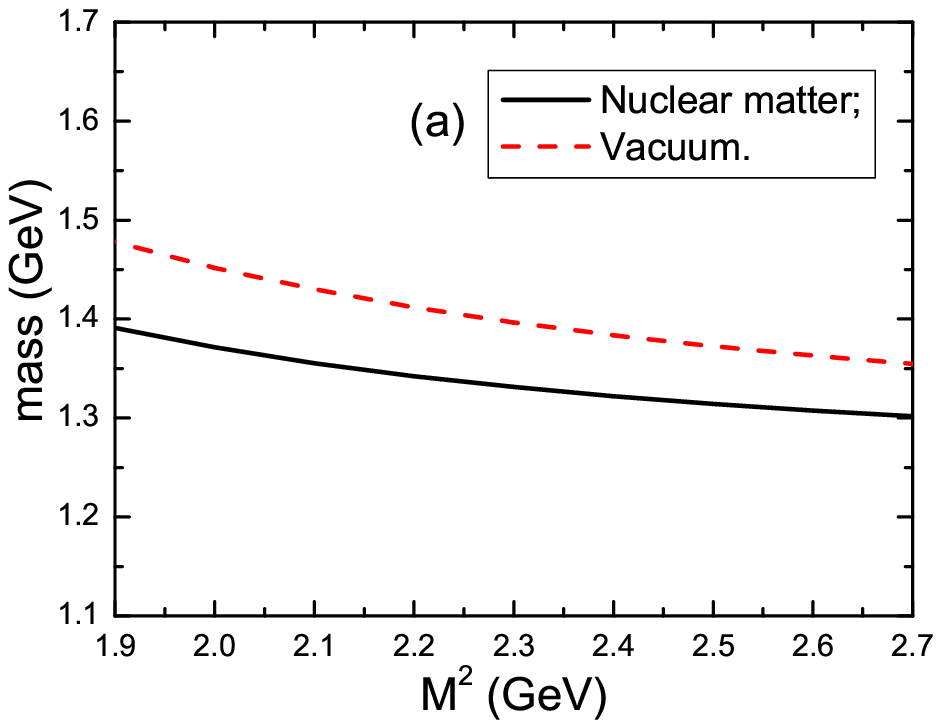}
 \includegraphics[totalheight=5cm,width=7cm]{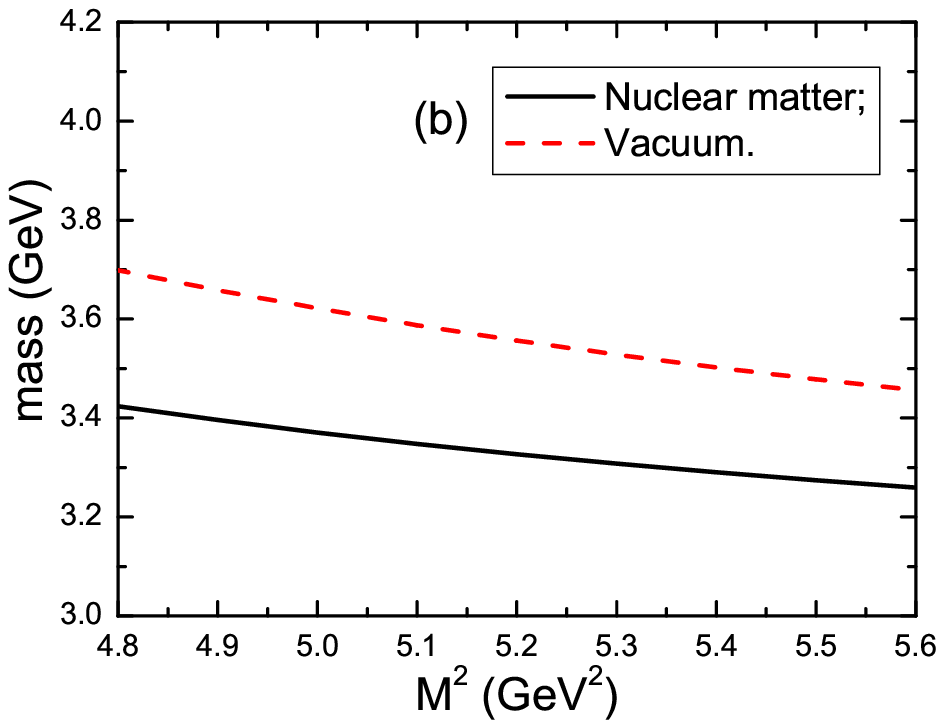}
 \caption{The  masses from the QCD sum rules in the vacuum and in the nuclear matter  versus the Borel parameter $M^2$, the (a) and (b) denote the
 $\Sigma_c$ and $\Sigma_b$ baryons, respectively. }
\end{figure}

\section{Conclusion}
In this article, we extend our previous work on the $\Lambda$-type heavy baryons $\Lambda_Q$  to   study the  properties  of the $\Sigma$-type
heavy baryons $\Sigma_Q$  in the nuclear matter using the QCD sum rules, and  derive three coupled QCD sum rules for
 the   masses $M_{\Sigma_Q}^*$, vector self-energies $\Sigma_v$ and pole residues $\lambda^*_{\Sigma_Q}$ in the nuclear matter, then  obtain the values   $M^*_{\Sigma_c}=1.33^{+0.06}_{-0.03}\,\rm{GeV}$, $M^*_{\Sigma_b}=3.33^{+0.09}_{-0.07}\,\rm{GeV}$, $\lambda_{\Sigma_c}^*=2.46^{+0.22}_{-0.16}\times 10^{-2}\,\rm{GeV}^3$, $\lambda_{\Sigma_b}^*=1.25^{+0.08}_{-0.04}\times 10^{-2}\,\rm{GeV}^3$, $\Sigma_{v}^{\Sigma_c}=0.446^{+0.035}_{-0.027}\,\rm{GeV}$,
  $\Sigma_{v}^{\Sigma_b}=0.776^{+0.042}_{-0.035}\,\rm{GeV}$. In the limit $\rho_N=0$, the predictions $M_{\Sigma_c}=1.40^{+0.08}_{-0.05}\,\rm{GeV}$ and  $M_{\Sigma_b}=3.56^{+0.14}_{-0.10}\,\rm{GeV}$ are much smaller than the experimental  data due to the unequal pole residues from different spinor structures $\gamma \cdot q$ and $1$.
 We normalize the masses $M_{\Sigma_c}=1.40\,\rm{GeV}$ and  $M_{\Sigma_b}=3.56\,\rm{GeV}$ from the  QCD sum rules in the vacuum to the experimental
  data, and obtain the mass-shifts in the nuclear matter $\delta M_{\Sigma_c}=-123\,\rm{MeV}$ and $\delta M_{\Sigma_b}=-375\,\rm{MeV}$, which can be confronted with the experimental data in the future.

\section*{Acknowledgments}
This  work is supported by National Natural Science Foundation,
Grant Number 11075053,  and the Fundamental
Research Funds for the Central Universities.

\end{document}